\include{graphicsx}

\documentclass[12pt,preprint]{aastex}

\slugcomment{To be published in the Feb 1, 2002 ApJ issue}

\shorttitle{Discovery of A Close Brown Dwarf Binary}
\shortauthors{Close et al.}

\begin{document}


\title{Discovery of a $0.15\arcsec$ Binary Brown Dwarf 2MASSJ
1426316+155701 With Gemini/Hokupa'a Adaptive Optics  }


\author{L. M. Close$^1$, D. Potter$^2$, W. Brandner$^3$, M. Lloyd-Hart$^1$, J. Liebert$^1$, A. Burrows$^1$, N. Siegler$^1$}

\email{lclose@as.arizona.edu}

\affil{$^1$Steward Observatory, University of Arizona, Tucson, AZ 85721}
\affil{$^2$Institute for Astronomy, University of Hawaii, Honolulu, HI}
\affil{$^3$European Southern Observatory, Garching, Germany}   



\begin{abstract} 

Use of the highly sensitive Hokupa'a curvature wavefront sensor has
allowed for the first time direct adaptive optics (AO) guiding on
brown dwarfs and VLM stars ($SpT=M7-L2$). An initial
survey of 9 such objects discovered one $0.15\arcsec$ binary (2MASSJ
1426316+155701). The companion is about half as bright as the primary
($\Delta K = 0.61\pm0.05$,$\Delta H = 0.70\pm0.05$) and has even
redder colors $H-K=0.59\pm0.14$ than the primary. The blended spectrum
of the binary has been previously determined to be M9.0. We modeled a
blend of an M8.5 template and a L1-L3 template reproducing a M9.0
spectrum in the case of $\Delta K = 0.61\pm0.05$,$\Delta H =
0.70\pm0.05$. These spectral types also match the observed H-K colors
of each star. Based the previously observed low space motion and
$H_{\alpha}$ activity we assign an age of $0.8^{+6.7}_{-0.3}
Gyr$. Utilizing this age range and the latest DUSTY models of the Lyon
group we assign a photometric distance of $18.8^{+1.44}_{-1.02} pc$
and masses of $M_{A}=0.074^{+0.005}_{-0.011} M_\odot$ and
$M_{B}=0.066^{+0.006}_{-0.015} M_\odot$. 
We therefore estimate a system separation of
$2.92_{+0.22}^{-0.16}AU$ and a period of $13.3_{+3.18}^{-1.51} yr$
respectively. Hence, 2M1426 is among the smallest separation brown
dwarf binaries resolved to date.


\end{abstract}

\keywords{instrumentation: adaptive optics --- binaries: general --- stars: evolution --- stars: formation
--- stars: individual (2MASSJ 1426316+155701) --- stars: low-mass,
    brown dwarfs}

\section{Introduction}

Since the discovery of Gl 229B by \cite{nak95} there has been intense
interest in the direct detection of brown dwarfs. To fully
understand this new population of sub-stellar objects it is critical to
characterize binary brown dwarfs (systems in which both components are
likely substellar). Brown dwarf binary systems are of critical importance since
they alone allow the masses and luminosities of brown dwarfs to be
measured directly (\cite{ken01}, \cite{lan01}). Without direct
dynamical mass measurements theoretical evolutionary models of brown
dwarfs (\cite{cha01},\cite{bur00}) cannot be calibrated. Therefore,
there is great interest in directly detecting close brown dwarf
binaries.

Moreover, the binary frequency of brown dwarfs is interesting in its
own right, since little is known about how common binary brown dwarf
systems are. It is not clear currently if the brown dwarf binary
distribution in mass ratio (q) and separation is similar to that of M
stars; in fact, there is emerging evidence that double brown
dwarfs (hereafter called binary brown dwarfs) tend to have mass ratios
close to unity and small separations (\cite{mar99}). However surveys
also suggest the binary frequency itself is similar to that of M
stars (\cite{rei01a}).

Despite the strong interest in such binary brown dwarfs, very few such
systems have been detected to date. Before HST/NICMOS ran out of
cryogen the first double L dwarf was detected (\cite{mar99}). A young
spectroscopic binary brown dwarf (PPL 15) was detected in the Pleiades
(\cite{bas99}), but this spectroscopic system is too close to get
separate luminosities for each companion. However, a large HST/NICMOS
imaging survey by \cite{mar00} of VLM dwarfs in the Pleiades failed to
detect any brown dwarf binaries with separations $>0.2\arcsec$
($>\sim27$ AU). Detections of closer binary systems were more
successful. The nearby object Gl 569B was resolved into a 0.1" (1 AU)
binary brown dwarf at KECK and the 6.5m MMT (\cite{mar99b};
\cite{ken01}; \cite{lan01}). Keck seeing-limited NIR 
imaging marginally resolved two more binary L stars (\cite{koe99}). A
survey with WFPC2 detected 4 more (3 newly discovered) close equal
magnitude binaries out of a sample of 20 L dwarfs
(\cite{rei01a}). Then \cite{pot01} guided on HD130948 with adaptive
optics and discovered a binary brown dwarf companion. Hence, the total
number of binary brown dwarfs known is just nine. In addition, of
these, only eight have luminosities known for each component, since
one is a spectroscopic binary.

In this paper we present a newly discovered binary brown dwarf 2MASSJ
1426316+155701 (hereafter 2M1426) which is a significant addition to
this list since it is the only example of a M8.5 primary with a L1-L3
secondary which is a large mass ratio compared to other brown dwarf
binaries. At the observed separation of $0.15\arcsec$, the
 2M1426 binary is the $4^{th}$
smallest separation resolved brown dwarf binary currently known, and so it
should likely play a significant role in the mass luminosity
calibration for brown dwarfs. Moreover, we show for the first time
that it possible to guide {\it directly} on a low mass star/brown
dwarf with adaptive optics on an 8m telescope, opening up a new
technique for diffraction-limited brown dwarf observations.


\section{Adaptive Optics Imaging with Gemini \& Hokupa'a}

Our approach to the detection of brown dwarf binaries is somewhat
different from that tried before. We wished to utilize both
diffraction-limited resolution and NIR ($1-2.5\micron$) imaging at the
wavelengths where brown dwarf spectral energy distributions peak. To
detect the closest binaries we utilized the largest aperture possible
(D$\sim 8$ m). Hence ground-based NIR imaging with adaptive optics
(AO) was a logical choice (cf. Close 2000 for a review on
AO). However, due to the extreme optical faintness of these brown
dwarfs ($V \sim 20$) most AO systems can not lock on such faint
targets. The exception are curvature based AO systems which employ
red-sensitive photon counting avalanche photodiodes (APDs) in their
wavefront sensors. Such a sensor can lock on a target as faint as V=20
as long as it is very red (V-I=4). Since nearby VLM stars/brown dwarfs
of spectral type M8-L1 have $V \sim 20$ and $V-I\sim 4$ we decided
such targets would be possible with a curvature AO system on an 8
meter telescope.
 
Recently, the University of Hawaii curvature AO system Hokupa'a
(Graves et al 1998; Close et al. 1998) was moved to the Gemini north
telescope and provided as a visitor instrument to the Gemini
community. Although Hokupa'a's 36 elements were too few to get very
high Strehl images on an 8m telescope, it was well suited to locking
onto nearby, faint, red M8-L1 stars and producing $0.1\arcsec$ images
(which is close to the $0.05\arcsec$ diffraction-limit at H band). We
decided to utilize this unique capability to survey the nearest
extreme M and L stars to characterize the nearby binary brown dwarf
population.

Here we report the results of our first observing run on June 20,
2001. We targeted VLM stars and brown dwarfs identified by
\cite{giz00}. In total, 9 brown dwarfs were observed over the first
half of the night. Typically we achieved $0.1-0.15\arcsec$ images of
$\sim10$ min total integration time in the H ($1.6\micron$)
band. Since the survey is on-going we will report in more detail about
all the objects observed in a more comprehensive paper in the near
future. In this paper we limit the discussion to the 2M1426 binary
observations.

One out of 9 of our targets (2M1426) was clearly a tight binary with
separation $\sim0.15\arcsec$. We observed the object dithering over 4
different positions on the QUIRC NIR 1024x1024 detector with
$0.0199\arcsec$/pix (\cite{hod96}). At each position we took 3x5s
exposures and 2x60s exposures. In addition, we took a series of 10x10s
K and K$^{\prime}$ images all in the same position.

\section{Reductions}

We have developed an AO data reduction pipeline in the IRAF language
which maximizes sensitivity and image resolution. The pipeline cross
correlates each image, then rotates each image so north is up and east
is left, then median combines the data with an average sigma clip
rejection at the $\pm2.5 \sigma$ level. By use of a cubic-spline
interpolator the script preserves image resolution to the $<0.02$ pixel
level. Next the script produces 2 final output images, one that
combines all the images taken, and another where only the sharpest 50\%
of the images are combined. The final images have a FOV of $30x30\arcsec$.

\begin{figure}
\includegraphics[angle=0,width=\columnwidth]{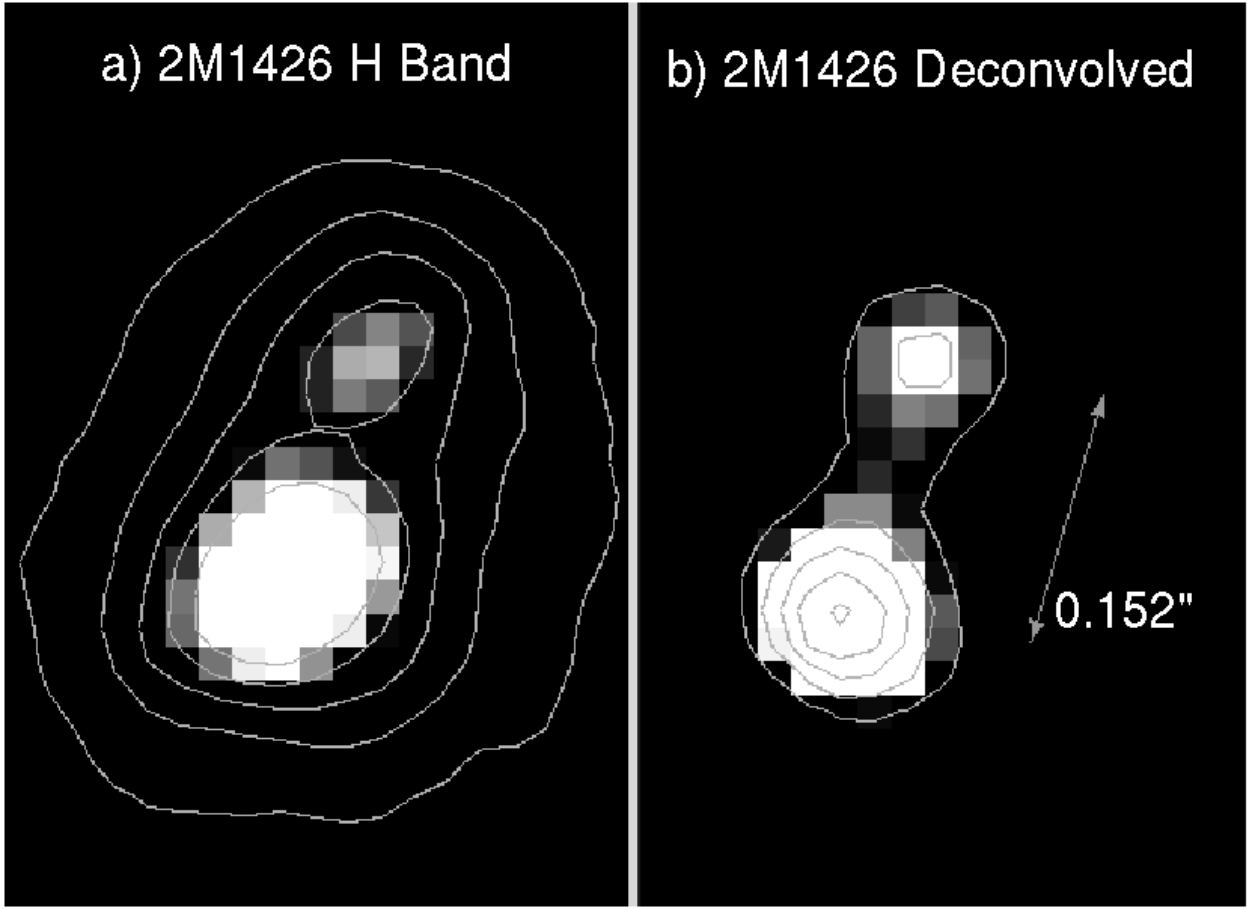}\caption{In figure (a) we see the 5s exposure 2MASSJ 1426316+155701 binary at an H band resolution of
$0.131\arcsec$. In Figure (b) we show the system after being LUCY
restored to $0.080\arcsec$ resolution. Only astrometry was derived
from the deconvolved images, photometry was derived from DAOPHOT PSF
fitting of the undeconvolved data. The pixels are
0.0199$\arcsec$/pix. The contours are linear at the 95, 75, 55, 35,
15\% levels.\label{fig1}} \end{figure}

This pipeline produced a final deep 480s H image (8x60s;
$FWHM=0.197\arcsec$), a final unsaturated 60s H image (12x5s;
$FWHM=0.131\arcsec$; see Figure
\ref{fig1}), and a unsaturated 50s K ($FWHM=0.197\arcsec$)
and K$^{\prime}$ (10x10s; $FWHM=0.197\arcsec$) image.

\placefigure{fig1} 

\section{Analysis}

In Table \ref{tbl-1} we present the analysis of the images taken of
2M1426. The photometry was based on DAOPHOT (\cite{ste87}) PSF fitting
photometry. The PSF used was the 12x5s unsaturated data from the next
brown dwarf observed after 2M1426. This PSF "star" had a similar
brightness (K=12.5) and late M8.5 spectral type and was observed at a
similar airmass. The flux ratio measured by DAOPHOT was normalized by the 
total flux of the binary in a $15\arcsec$ aperture. The resulting
magnitudes are listed in Table \ref{tbl-1}. 

The platescale and orientation of QUIRC was determined from a short
exposure of the Trapezium cluster in Orion and compared to published
positions as in \cite{sim99}. From these observations a platescale of
$ 0.0199\arcsec$/pix and an orientation of the Y-axis (0.3 degrees E
of north) was determined.

The astrometry was based on lightly (100 iterations) LUCY deconvolved
data from the 12x5s unsaturated H image (see Figure
\ref{fig1}). For the
deconvolution we used the same PSF as above. A good
deconvolution result was obtained with a restored resolution of
$FWHM=0.080\arcsec$. This high resolution allowed for excellent
astrometric accuracy to be achieved by centroiding on the well separated
stars. These astrometric measurements were also
checked against the PSF fitting photometry of DAOPHOT and were found
to be in excellent agreement (DAOPHOT gave $separation=0.155\arcsec$,
$PA=343.5^{o}$).

\placetable{tbl-1}

\placetable{tbl-2}

\section{Discussion}

\subsection {Is the companion physically related to the 2M1426 system?}

We believe there is a very high probability that the companion is
physically associated with the 2M1426 system. This is likely due to
the small space density of very red ($H-K=0.59$) background objects in
the field. Furthermore, in the 9000 square arcsecs already surveyed
we have not detected a similarly reddened background object in any of
the fields. Therefore, we estimate the probability of a chance
projection of such a red object within $0.15\arcsec$ of the primary is
$<1x10^{-5}$. We conclude that this very red, cool object is
physically related to the 2M1426 primary and hereafter refer to it as
2M1426B.

\subsection {What are the spectral types of the components?}

We do not have spatially resolved spectra of both components; consequently we
can only try to "blend" spectra of a hotter and cooler star while
preserving the observed $\Delta K=0.61$ and $\Delta H=0.70$. In
\cite{ken01} it was found that a M8.5 and a L1-L3 spectrum could be
blended together (with $\Delta K=0.61$) to produce a M9 ``blended''
spectrum. Since \cite{giz00} observed 2M1426 to have a blended M9.0
spectrum we can reasonably assume that the components may have spectral
types of M8.5 and L1-L3. Hence we estimate (until a proper
spatially resolved spectrum is obtained of both components) that
2M1426A is similar to a M8.5 and 2M1426B is similar to a L1-L3
spectrum. Furthermore, the $H-K=0.47$ color of the primary is
consistent with a M8.5 SpT. Similarly, the $H-K=0.59$ color of the
companion is consistent with a L1-L3 SpT \cite{rei01b}. So we will
assume a M8.5 SpT for the primary and a L1-L3 SpT for the companion
until proper spatially resolved spectra are obtained and the true
spectral types determined.

\subsection {What is the distance to the 2M1426 system?}

Unfortunately, there is no published parallax to the 2M1426 system. We
can estimate, however, the distance based on the spectral types and a
range of possible ages for the system. To do this, calibrated
theoretical evolutionary tracks are required for objects in the
temperature range 2300-1800 K. Recently such a calibration has been
performed by two groups using dynamical measurements of the Gl569B
brown dwarf binary. From the dynamical mass measurements of the GL569B
binary brown dwarf (\cite{ken01}, and \cite{lan01}) it was found that
the \cite{cha01} and \cite{bur00} evolutionary models were in reasonably
good agreement with observation.  Based on the latest \cite{cha01} DUSTY
models (see Figure
\ref{fig2}) we
find that even if the age of 2M1426 was as old as 7.5Gyr the closest it
could be is 17.8 pc based on the observed $K_{A}=12.16$ and
$SpT_{A}$=M8.5. On the other hand, if 2M14216 is as young as 0.5 Gyr
it could be as far as 20.3 pc with $K_{A}=12.16$ and
$SpT_{A}$=M8.5. Since it is highly likely that the age of 2M1426 is
between 0.5-7.5 Gyr, we adopt a distance range of 17.8-20.3 pc.

\placefigure{fig2}

\subsection {What is the age of the 2M1426 system?}

Estimating the age for the 2M1426 system is difficult since there are
no Li measurements yet made. The age could be anywhere from 0.5-7.5
Gyr. However, the low proper motion observed by \cite{giz00} for
the system ($\mu RA=0.108\arcsec/yr$, $\mu DEC=-0.056\arcsec/yr$)
gives $V_{tan}=12$ km/s. Unfortunately the low resolution of the
spectrum obtained by \cite{giz00} does not allow the radial velocity
(and hence full 3-dimensional space motion) to be calculated at this
time. However, such a low $V_{tan}=12$ km/s velocity makes 2M1426 the
lowest velocity M8-M9 in the entire survey of
\cite{giz00}. This strongly suggests a young age since the system has
not developed a significant random velocity like the other older M8-M9
stars in the survey. Moreover, the low 1.2 Angstrom EW of $H_{\alpha}$
may also suggest a youthful age for such a late spectral type (where
$H_{\alpha}$ appears more strongly in older systems;
\cite{giz00}). Indeed, 2M1426 occupies a uniquely young position as
the lowest $H_{\alpha}$ and lowest $V_{tan}$ of any of the 14 M8-M9
objects in the \cite{giz00} survey. Therefore, we assign a young age of
$0.8^{+6.7}_{-0.3}$ Gyr to this system, but leave large error bars.

\subsection {How does 2M1426 compare to the other binary dwarfs known?}

As was mentioned in the introduction there are 8 binary brown dwarfs
(published at this time) with resolved components
(i.e. non-spectroscopic). A listing of these in order of separation
can be seen in Table \ref{tbl-3}.

\placetable{tbl-3}

From Table \ref{tbl-3} we see that, other than the very close (10pc)
system GL569B, the next 4 widest systems all have estimated periods in
the range 11-16 yr. These periods in Table \ref{tbl-3} are crudely
estimated by assuming face-on circular orbits. However, even such
crude estimates do allow us to see that 2M1426 is in the top 5 of the
shortest period brown dwarf binaries currently known.  This is of
great importance since it is only those systems with separations less
than 4 AU that will have periods less than 25 yrs. Since a good
fraction of a period is required to calculate the dynamical mass, such
objects are of great importance. The shortest period resolved brown
dwarf binary (Gl 569B) has already had its orbit calculated and the
first zero-point to the theoretical tracks has been applied
(\cite{ken01};
\cite{lan01}). However, one measurement does not calibrate the 3-dimensional Mass-Luminosity-Age tracks (see Figure
\ref{fig2}), and so dynamical
masses and luminosities are required for more of the systems in Table
\ref{tbl-3}. Hence, the short period binary 2M1426 makes a significant addition to the known brown dwarf binaries.

\subsection {The masses of the components}

Unfortunately no Li observations exist yet for the system, so we must
use the models in Figure
\ref{fig2} to estimate the masses. We
can estimate the masses of the components based on the age range of
0.5-7.5 Gyr. The models suggest a mass range of $M_{A}= 0.063-0.079
M_{\odot}$ and $M_{B}=0.051-0.072 M_{\odot}$ for our spectral types of
M8.5 and L1 and the age range of 0.5-7.5 Gyr, respectively. At the
older ages ($>1 Gyr$) the primary may be on the stellar/substellar
boundary, but the substellar nature ($M_{B}<0.075 M_{\odot}$) of the
companion is certain even in the unlikely age 7.5 Gyr, according to
the models.

\subsection {Future observations}

   Future observations of this binary should determine if there is
   still Li present in its spectrum. The best Li observation would
   spatially resolve both components in the visible
   spectrum. Currently there is no visible wavelength AO systems
   capable of guiding on a $V\sim 20$ source and obtaining resolutions
   better than $0.1 \arcsec$ \citep{clo00}. Therefore, we will have
   to carry out such observations from space with HST/STIS
   perhaps. Parallax measurements should also be obtained in the near
   future. Within the next few years one should also be able to
   measure the masses of both components as they complete a
   significant fraction of their orbit. Hence, further observations of
   2M1426 are a high priority for the calibration of the brown dwarf
   mass-age-luminosity relation.

\acknowledgements

\begin{figure}
\includegraphics[angle=270,width=\columnwidth]{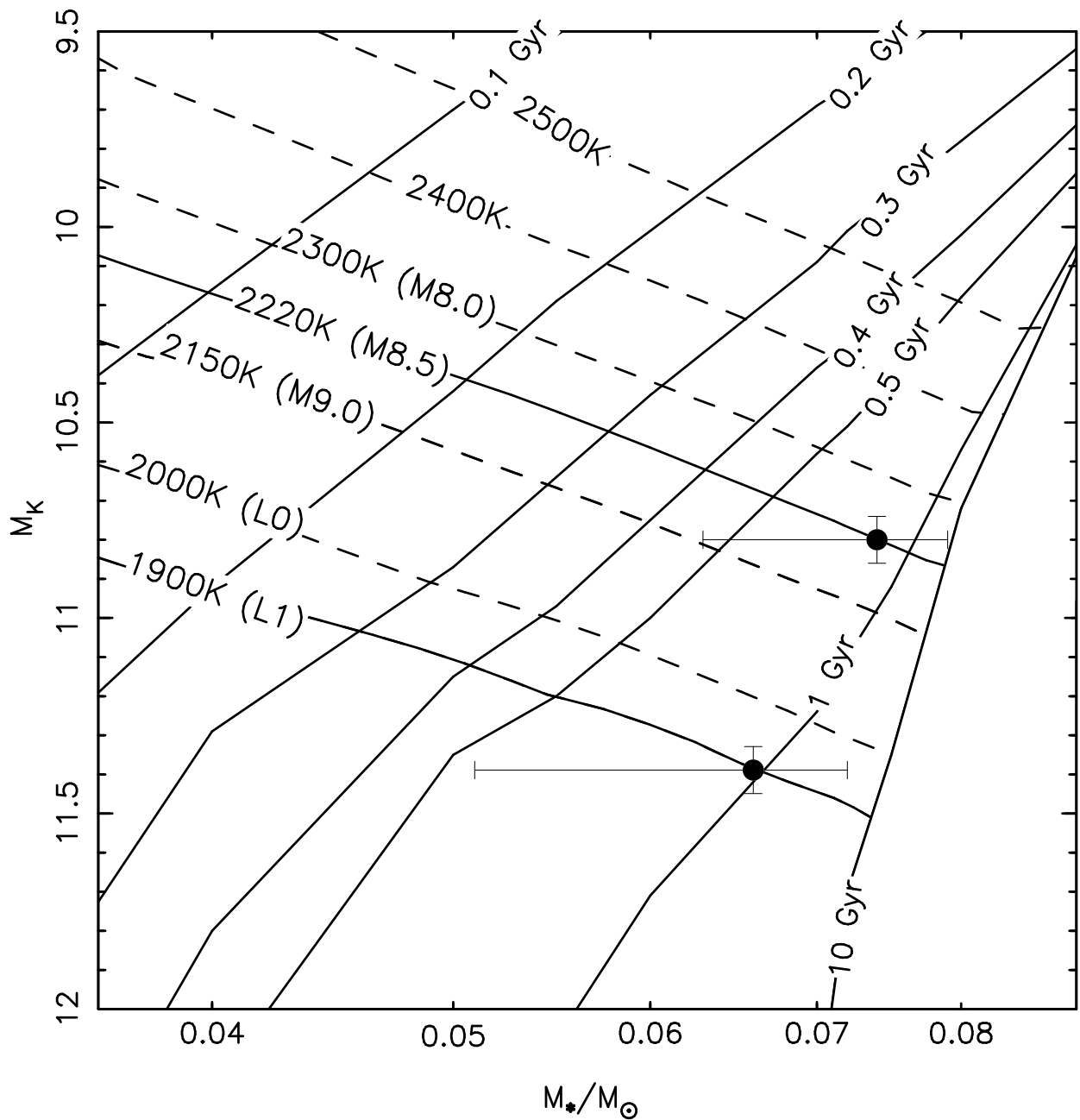}\caption{The latest \cite{cha01} DUSTY evolutionary models, and the temperature
scale of \cite{leg01}. The locations of the 2 components of 2M1426 are
indicated by the solid circles. Note the good agreement of the secondary's
position with the 0.8 Gyr isochrone and the 1900K ($\sim$L1)
isotherm. The large error bars are due to the errors in the uncertain age of the system $0.8^{+6.7}_{-0.3}Gyr$. \label{fig2}} \end{figure}

We thank Isabelle Baraffe for supplying us with theoretical models
computed by herself, France Allard, Gilles Chabrier and Peter
Hauschildt. Mathew Kenworthy helped produce Figure
\ref{fig2}. LMC acknowledges support by the AFOSR under
F49620-00-1-0294. Observations in this paper were carried out with the
Gemini North Telescope. The Hokupa'a AO observations were supported by
the University of Hawaii AO group (D. Potter, O. Guyon, P. Badouz and
A. Stockton). Support for Hokupa'a comes from the National Science
Foundation. Olivier Guyon kindly provided an algorithm for the
re-rotation of Quirc images. We would also like to send a big {\it
mahalo nui} to the Gemini operations staff (especially Francois Rigaut, 
Simon Chan 
\& John Hamilton) for a flawless night. As well Steve Ridgway (NOAO)
was a great help in acquiring the data at Gemini.

These results were based on observations obtained at the Gemini
Observatory, which is operated by the Association of Universities for
Research in Astronomy, Inc., under a cooperative agreement with the
NSF on behalf of the Gemini partnership: the National Science
Foundation (United States), the Particle Physics and Astronomy
Research Council (United Kingdom), the National Research Council
(Canada), CONICYT (Chile), the Australian Research Council
(Australia), CNPq (Brazil) and CONICET (Argentina).





\clearpage
\begin{deluxetable}{llll}
\tabletypesize{\scriptsize}
\tablecaption{The 2M1426 System June 20, 2001 \label{tbl-1}}
\tablewidth{0pt}
\tablehead{
\colhead{Observed} &
\colhead{Value} &
\colhead{Calculated} &
\colhead{Value} \\
\colhead{Parameter} &
\colhead{} &
\colhead{Parameter} &
\colhead{} 
}
\startdata
$\Delta K$  & $0.61\pm0.05$ & Age        & $0.8^{+6.7}_{-0.3}Gyr$       \\
$\Delta K^{\prime}$ & $0.65\pm0.10$ & Distance\tablenotemark{a}   & $18.8^{+1.44}_{-1.02} pc$ \\
$\Delta H$  & $0.70\pm0.05$ & Separation & $2.92_{+0.22}^{-0.16}AU$ \\
Ang. Sep. & $0.152 \pm 0.005\arcsec$ & $M_{tot}$ & $0.140^{+0.011}_{-0.026} M_{\odot}$ \\
PA     & $344.05\pm0.5^{o}$ & Est. Period & $13.3_{+3.18}^{-1.51} yr$ \\
\enddata
\tablenotetext{a}{photometric distances from the models of \cite{cha01}}
\end{deluxetable}

\clearpage
\begin{deluxetable}{lll}
\tabletypesize{\scriptsize}
\tablecaption{Summary of the 2M1426 A \& B components \label{tbl-2}}
\tablewidth{0pt}
\tablehead{
\colhead{Parameter} &
\colhead{2M1426A} &
\colhead{2M1426B}
}
\startdata
H mag   & $12.63\pm 0.05$ & $13.34\pm 0.10$ \\
Ks mag  & $12.20\pm 0.07$ & $12.80\pm 0.14$ \\
K mag   & $12.16\pm 0.05$ & $12.75\pm 0.10$ \\
H-K color & $0.47\pm 0.08$ & $0.59\pm 0.14$  \\
Est. SpT & M8.5            & L1-L3              \\
Mass & $0.074^{+0.005}_{-0.011} M_\odot$ & $0.066^{+0.006}_{-0.015} M_\odot$ \\
\enddata
\end{deluxetable}

\clearpage
\begin{deluxetable}{llll}
\tabletypesize{\scriptsize}
\tablecaption{Published Resolved\tablenotemark{a} Brown Dwarf Binaries \label{tbl-3}}
\tablewidth{0pt}
\tablehead{
\colhead{Name} &
\colhead{Separation} &
\colhead{Est. Period\tablenotemark{b}} &
\colhead{Ref.\tablenotemark{c}} \\
\colhead{} &
\colhead{AU} &
\colhead{yr} &
\colhead{} 
}
\startdata
Gl 569B\tablenotemark{d}  &  1.0  & 3 & 1,2 \\
HD130948B\tablenotemark{d} &  2.5  & 11 & 3  \\
2M0746   &  2.7  & 12 & 4  \\
2M1426   &  3.1-2.7\tablenotemark{e} & 16-12\tablenotemark{e} & this paper \\
2M0920   &  3.2  & 16  & 4 \\
2M0850   &  4.4  & 28  & 4 \\
DENIS 1228 & 4.9 & 33 &  5 \\
2M1146	& 7.6 & 63 & 6 \\
DENIS 0205 & 9.2 & 72 & 6 \\
\enddata
\tablenotetext{a}{resolved binaries have measured fluxes for each component, the one spectroscopic binary PPL 15 (\cite{bas99}) is excluded}
\tablenotetext{b}{This ``period'' is simply an estimate assuming a face-on circular orbit}
\tablenotetext{c}{REFERENCES--(1) \cite{ken01}; (2) \cite{lan01}; (3) \cite{pot01}; (4)\cite{rei01a}; (5) \cite{mar99}; (6) \cite{koe99}}
\tablenotetext{d}{Gl 569B and HD130948B are binary brown dwarfs that orbit normal Stars, for AO observations these bright primary stars were guided on, not the brown dwarfs}
\tablenotetext{e}{values for age range 0.5-7.5 Gyr respectively}
\end{deluxetable}

\clearpage

\clearpage

\end{document}